\def\year{2018}\relax
\documentclass[letterpaper]{article} %DO NOT CHANGE THIS
\usepackage{aaai18}  %Required
\usepackage{times}  %Required
\usepackage{helvet}  %Required
\usepackage{courier}  %Required
\usepackage{url}  %Required
\usepackage{graphicx}  %Required

\usepackage{amsmath}
\usepackage{latexsym}
\usepackage{subfig}

\begin{document}
\title{A Hierarchical Recurrent Neural Network for Symbolic Melody Generation}
\author{Jian Wu,
        Changran Hu,
        Yulong Wang,
        Xiaolin Hu,
        and Jun Zhu}

\maketitle

\begin{abstract}
In recent years, neural networks have been used to generate symbolic melodies.
However, the long-term structure in the melody has posed great difficulty for designing a good model.
In this paper, we present a hierarchical recurrent neural network for melody generation, which consists of three Long-Short-Term-Memory (LSTM) subnetworks working in a coarse-to-fine manner along time.
Specifically, the three subnetworks generate bar profiles, beat profiles and notes in turn, and the output of the high-level subnetworks are fed into the low-level subnetworks, serving as guidance for generating the finer time-scale melody components in low-level subnetworks.
Two human behavior experiments demonstrate the advantage of this structure over the single-layer LSTM which attempts to learn all hidden structures in melodies.
% In the third human behavior experiment, subjects are asked to judge whether the generated melody is composed by human or computer.
% The results show that 33.69\% of the generated melodies are wrongly classified as human composed.
% In addition, only 74.95\% of human composed melodies are correctly classified, which indicates that the model has achieved a good result.
Compared with the state-of-the-art models MidiNet~\cite{yang2017midinet} and MusicVAE~\cite{pmlr-v80-roberts18a}, the hierarchical recurrent neural network produces better melodies evaluated by humans. 
\end{abstract}

\section{Introduction}
Automatic music generation using neural networks has attracted much attention.
There are two classes of music generation approaches, symbolic music generation~\cite{hadjeres2016deepbach}\cite{magenta2016}\cite{yang2017midinet} and audio music generation~\cite{oord2016wavenet}\cite{mehri2016samplernn}.
In this study, we focus on symbolic melody generation, which requires learning from sheet music.

Many music genres such as pop music consist of melody and harmony.
Since usually beautiful harmonies can be ensured by using legitimate chord progressions which have been summarized by musicians, we only focus on melody generation, similar to some recent studies~\cite{magenta2016}\cite{yang2017midinet}\cite{colombo2017deep}\cite{pmlr-v80-roberts18a}.
This greatly simplifies the melody generation problem.

Melody is a linear succession of musical notes along time.
It has both short time scale such as notes and long time scale such as phrases and movements,
which makes the melody generation a challenging task.
Existing methods generate pitches and rhythm simultaneously \cite{magenta2016} or sequentially \cite{chu2016song} using Recurrent Neural Networks~(RNNs), but they usually work on the note scale without explicitly modeling the larger time-scale components such as rhythmic patterns.
It is difficult for them to learn long-term dependency or structure in melody.

Theoretically, an RNN can learn the temporal structure of any length in the input sequence, but in reality, as the sequence gets longer it is very hard to learn long-term structure.
Different RNNs have different learning capability, e.g., LSTM~\cite{hochreiter1997long} performs much better than the simple Elman network.
But any model has a limit for the length of learnable structure, and this limit depends on the complexity of the sequence to be learned.
To enhance the learning capability of an RNN, one approach is to invent a new structure.
In this work we take another approach: increase the granularity of the input.
Since each symbol in the sequence corresponds to longer segment than the original representation, the same model would learn longer temporal structure.

To implement this idea, we propose a Hierarchical Recurrent Neural Network (HRNN) for learning melody. It consists of three LSTM-based sequence generators --- Bar Layer, Beat Layer and Note Layer.
% These generators are trained on bar sequences, beat sequences and note sequences extracted from sheet music, respectively.
The Bar Layer and Beat Layer are trained to generate bar profiles and beat profiles, which are designed to represent the high-level temporal features of melody.
The Note Layer is trained to generate melody conditioned on the bar profile sequence and beat profile sequence output by the Bar Layer and Beat Layer.
By learning on different time scales, the HRNN can grasp the general regular patterns of human composed melodies in different granularities, and generate melody with realistic long-term structures. This method follows the general idea of granular computing~\cite{bargiela2012granular}, in which different resolutions of knowledge or information is extracted and represented for problem solving.
With the shorter profile sequences to guide the generation of note sequence, the difficulty of generating note sequence with well-organized structure is alleviated.
% If the maximum length of long-term structure generated without profiles as condition is $L$, then the maximum length of long-term structure generated with profiles is $16L$, in theoretically.

\section{Related Work}

\subsection{Melody Generation with Neural Networks}
There is a long history of generating melody with RNNs.
A recurrent autopredictive connectionist network called CONCERT is used to compose music~\cite{mozer1994neural}.
With a set of composition rules as constraints to evaluate melodies, an evolving neural network is employed to create melodies~\cite{chen2001creating}.
As an important form of RNN, LSTM~\cite{hochreiter1997long} is used to capture the global music structure and improve the quality of the generated music~\cite{eck2002first}.
% Eck and Schmidhuber use LSTM \cite{hochreiter1997long} to capture the global music structure and improve the quality of the generated music~\cite{eck2002first}.
Boulanger-Lewandowski, Bengio, and Vincent explore complex polyphonic music generation with an RNN-RBM model~\cite{boulanger2012modeling}.
Lookback RNN and Attention RNN are proposed to tackle the problem of creating melody's long-term structure~\cite{magenta2016}.
The Lookback RNN introduces a handcrafted lookback feature that makes the model repeat sequences easier while the Attention RNN leverages an attention mechanism to learn longer-term structures.
Inspired by convolution, two variants of RNN are employed to attain transposition invariance~\cite{johnson2017generating}.
%Some approaches have been proposed to learn the complex temporal structures of melody.
%Lookback feature \cite{magenta2016} allows the model to more easily recognize patterns in a certain time length and generate melodies that wander less with some simple musical structure.
To model the relation between rhythm and melody flow, a melody is divided into pitch sequence and duration sequence and these two sequences are processed in parallel~\cite{colombo2016algorithmic}.
This approach is further extended in \cite{colombo2017deep}.
A hierarchical VAE is employed to learn the distribution of melody pieces in~\cite{pmlr-v80-roberts18a}, the decoder of which is similar to our model. 
The major difference is that the higher layer of its decoder uses the automatically learned representation of bars, while our higher layers use predefined representation of bars and beats which makes the learning problem easier.
Generative Adversarial Networks~(GANs) have also been used to generate melodies. For example,
RNN-based GAN~\cite{mogren2016c} and CNN-based GAN~\cite{yang2017midinet} are employed to generate melodies, respectively.
However, the generated melodies also lack realistic long-term structures.

Some models are proposed to generate multi-track music. 
A 4-layer LSTM is employed to produce the key, press, chord and drum of pop music seperately~\cite{chu2016song}.
With pseudo-Gibbs sampling, a model can generate highly convincing chorales in the style of Bach~\cite{colombo2017deep}.
Three GANs for symbolic-domain multi-track music generation were proposed~\cite{DBLP:conf/aaai/DongHYY18}.
An end-to-end melody and arrangement generation framework XiaoIce Band was proposed to generate a melody track with accompany tracks with RNN~\cite{Zhu:2018:XBM:3219819.3220105}.

\subsection{Hierarchical and Multiple Time Scales Networks}
The idea of hierarchical or multiple time scales has been used in neural network design, especially in the area of natural language processing.
The Multiple Timescale Recurrent Neural Network~(MTRNN) realizes the self-organization of a functional hierarchy with two types of neurons ``fast'' unit and ``slow'' unit~\cite{yamashita2008emergence}.
Then it is shown that the MTRNN can acquire the capabilities to recognize, generate, and correct sentences in a hierarchical way: characters grouped into words, and words into sentences~\cite{hinoshita2011emergence}.
An LSTM auto-encoder is trained to preserve and reconstruct paragraphs by hierarchically building embeddings of words, sentences and paragraphs~\cite{li2015hierarchical}.
To process inputs at multiple time scales, the Clockwork RNN is proposed, which partitions the hidden layers of RNN into separate modules~\cite{koutnik2014clockwork}.
% Koutnik et al. propose Clockwork RNN that partition the hidden layer of RNN into separate modules to process inputs at multiple time scales~\cite{koutnik2014clockwork}.
Different from the Clockwork RNN, we integrate the prior knowledge of music in constructing the hierarchical model and feed multiple time scales of features to different layers.

\section{Music Concepts and Representation}
\label{Sec:BMT}
We first briefly introduce some basic music concepts and their properties to familiarize the readers who do not have a music background, then explain how the concepts are represented in the model.
% We first introduce the melody representation and then the handcrafted rhythmic profiles for HRNN.
% For readers who do not have a music background, some basic music concepts are introduced in the
% \textbf{Supplementary Text and Fig.~S1}.

\subsection{Basic Music Concepts}

\begin{figure}
\centering
\includegraphics[width=0.48\textwidth]{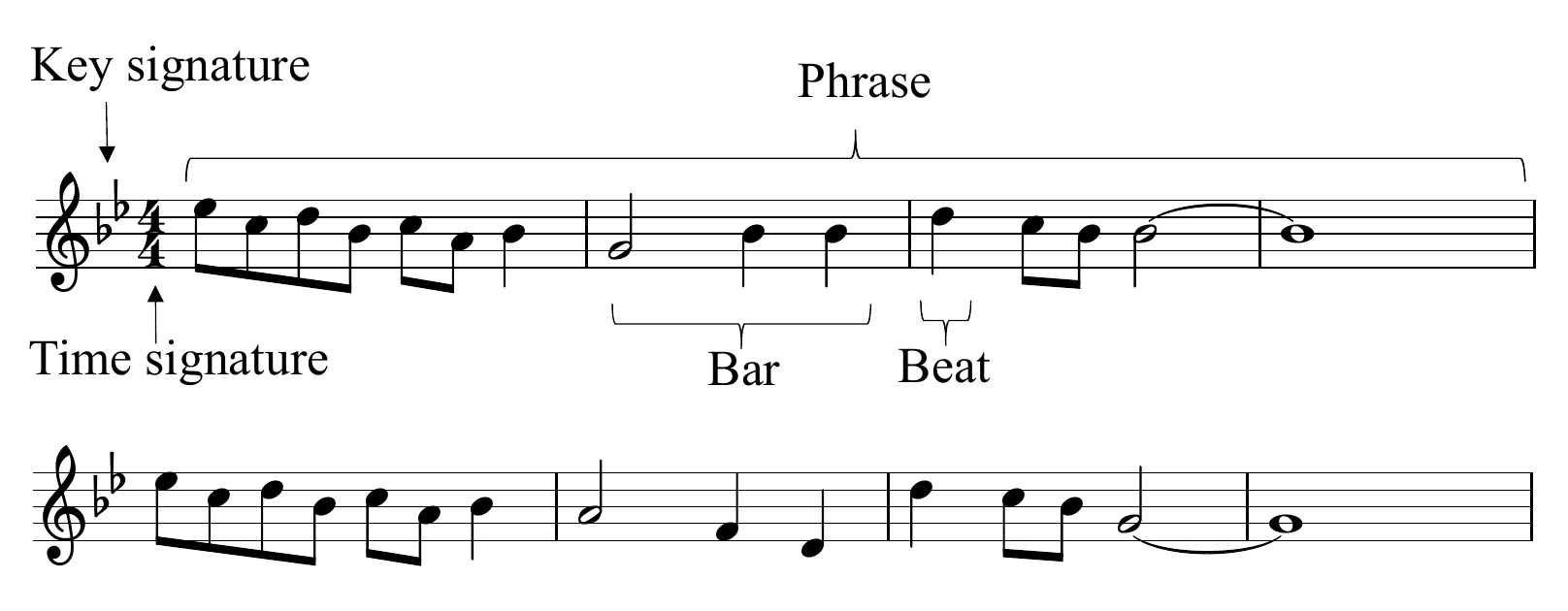}
\caption{A typical form of melody. The time signature of this musical piece is 4/4. The numerator means a bar contains 4 beats, and the denominator means the time length of 1 beat is a quarter note.}
\label{Fig:Melody}
\end{figure}

As shown in Fig.~\ref{Fig:Melody}, melody, often known as tune, voice, or line, is a linear succession of musical notes, and each note represents the pitch and duration of a sound. Several combined notes form a beat that determines the rhythm based on which listeners would tap their fingers when listening to music.
A bar contains a certain number of beats in each musical piece.
Time signature~(e.g., 3/4) specifies which note value is to be given in each beat by the denominator and the number of beats in each bar by the numerator.
Each musical piece has a key chosen from 12 notes in an octave.
Key signature, such as C$\sharp$ or B$\flat$, designates which key the current musical piece is.
The musical piece can be transformed to different keys while maintaining the general tone structure.
Therefore we can transpose all of the musical pieces to key C, while maintaining the relative relationship between notes.
Shifting all musical pieces to the same key makes it easier for the model to learn the relative relationship between notes.
The generated pieces can be transposed to any key.

\begin{figure*}
\centering
\includegraphics[width=\textwidth]{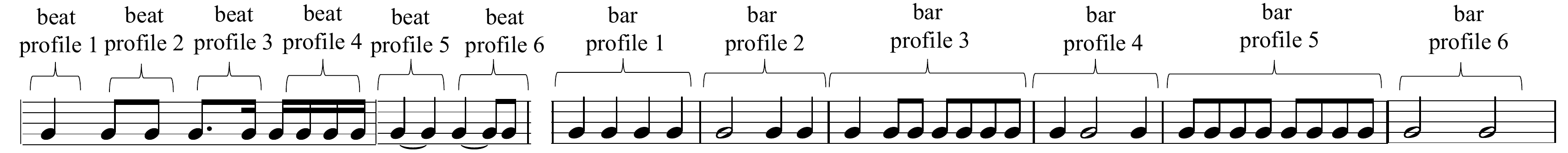}
\caption{
Samples of beat profiles and bar profiles.
Here we use notes with same pitch to illustrate rhythm in beat and bar.
The rhythm represented by beat profile 5 and 6 are related with the rhythm of the previous beat so they are shown with two beats where the first beats are all quarter notes.
}
\label{Fig:Profiles}
\end{figure*}

\begin{figure}
\centering
\includegraphics[width=0.8\linewidth]{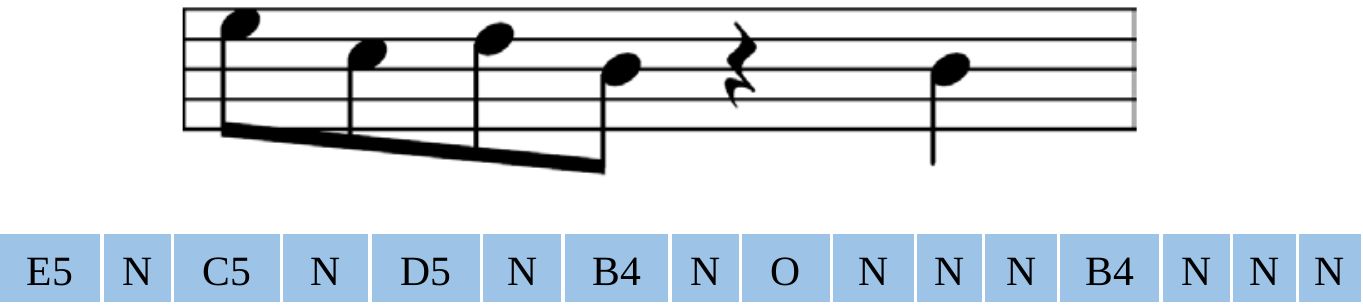}
\caption{An example of melody representation. \textbf{Top}: A melody with the length of one bar. \textbf{Bottom}: Representation of the melody, in which the N means a no-event and the O means a note-off event. Since the fourth note is not immediately followed by any note, a note-off event is necessary here.}
\label{Fig:Representation}
\end{figure}

\begin{figure*}[htpb]
\centering
\includegraphics[width=\textwidth]{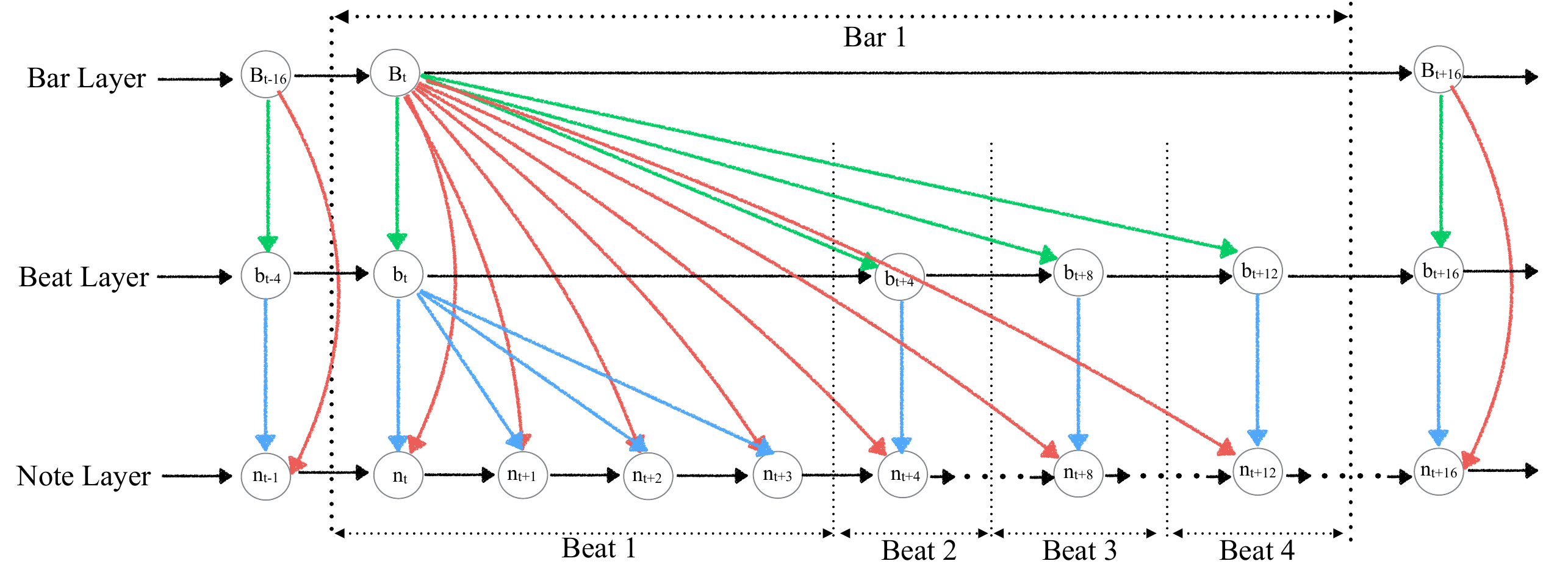}
\caption{Architecture of HRNN. From top to bottom are Bar Layer, Beat Layer and Note Layer respectively. Inner layer connections along time are shown with black lines. Connections between layers are shown with green lines, blue lines and red lines. }
\label{Fig:Arch}
\end{figure*}

\subsection{Melody Representation}
\label{Sec:MelodyRepresentation}
To simplify the problem, we only chose musical pieces with the time signature 4/4. This is a widely-used time signature.
% One property of melody is that its pitches are usually limited in C3 to C6.
According to the statistics on the Wikifonia dataset described in Section~\ref{Sec:Exp}, about 99.83\% of notes have pitches between C2 and C5.
Thus, all notes are octave-shifted to this range. Then there are 36 options for a pitch of a note (3 octaves and each octave has 12 notes).
To represent duration, we use event messages in the Midi standard.
When a note is pressed, a note-on event with the corresponding pitch happens; and when the note is released, a note-off event happens.
% For a monophonic melody, a new note-on event also indicates the previous note's note-off event.
For a monophonic melody, if two notes are adjacent, the note-on event of the latter indicates the note-off event of the former, and the note-off event of the former is therefore not needed.
In this study, every bar was discretized into 16 time steps.
At every time step, there are 38 kinds of events (36 note-on events, one note-off event and one no-event), which are exclusive.
One example is shown in Fig.~\ref{Fig:Representation}.
In this way, note-on events mainly determine the pitches in the melody and no-events mainly determine the rhythm as they determine the duration of the notes.
So a 38-dimensional one-hot vector is used to represent the melody at every time step.

\subsection{Rhythmic Patterns and Profiles}

Rhythmic patterns are successions of durations of notes which occur periodically in a musical piece.
% Generally, a musical piece is considered as a unity mainly because of the repeated rhythmic patterns.
% Phrases in the same musical piece may have similar rhythmic patterns, where the specific pitch of notes can be different.
It is a concept on a larger time scale than the note scale and is important for melodies' long-term structure.
Notice that in this model we do not encode the melody flow because it is hard to find an appropriate high-level representation of it.

% In order to learn the rhythmic patterns more efficiently, we design two features named \textit{beat profile} and \textit{bar profile} which are high-level representations of a whole bar and beat, respectively.
Two features named \textit{beat profile} and \textit{bar profile} are designed, which are high-level representations of a whole bar and beat, respectively.
Compared with individual notes, the two profiles provide coarser representations of the melody.
To construct the beat profile set, all melodies are cut into melody clips with a width of one beat and binarized at each time step with $1$ for an event~(note-on events and note-off event) and $0$ for no-event at this step.
Then we cluster all these melody clips into several clusters via the K-Means algorithm and use the cluster centers as our beat profiles.
Given a one beat melody piece, we can binarize it in the same manner and choose the closest beat profile as its representation.
The computation of bar profile is similar, except that the width of melody clip is changed to one bar.
Based on the well-known elbow method, the numbers of clusters for beat profiles and bar profiles are set to be 8 and 16 respectively.
In Fig.~\ref{Fig:Profiles}, some frequently appeared profiles are shown with notes.

\section{Hierarchical RNN for Melody Generation}
\label{Sec:Methods}

\subsection{Model Architecture}\label{Method:HMG}
HRNN consists of three event sequence generators: Bar Layer, Beat Layer and Note Layer, as illustrated in Fig.~\ref{Fig:Arch}.
These layers are used to generate bar profile sequence, beat profile sequence and note sequence, respectively.

The lower-level generators generate sequences conditioned on the sequence output by the higher-level generators.
So to generate a melody, one needs to first generate a bar profile sequence and a beat profile sequence in turn.
Suppose that we want to generate a melody piece with the length of one bar, which is represented as $n_t, ..., n_{t+15}$(see Fig.~\ref{Fig:Arch}).
First, the Bar Layer generates a bar profile $B_{t}$ with the last bar profile $B_{t-16}$ as input.
Then the Beat Layer generates 4 beat profiles $b_t, ..., b_{t+12}$ with  $b_{t-4}$ as input conditioned on the bar profile $B_t$.
To generate the notes $n_t, n_{t+1},...n_{t+3}$, the Note Layer is conditioned on both $B_t$ and $b_t$;
to generate the notes $n_{t+4}, ..., n_{t+7}$, the Note Layer is conditioned on both $B_t$ and $b_{t+4}$; and so on.
In this way, each bar profile is a condition for the 16 generated notes and each beat profile is a condition for the 4 generated notes.

All of the three layers use LSTM but the time scales of the input are different.
Theoretically, the Beat Layer and Bar Layer can learn 4 and 16 times longer temporal structure than the Note Layer, respectively.
Note that it is difficult to quantify the length of temporal structure learned in a model, since ``temporal structure'' is an abstract concept and its characterization is still an open problem.
We could only probe the difference in length produced by different models indirectly by measuring the quality of the generated sequences using behavior experiments (see Section 5).

To explicitly help RNN memorize recent events and potentially repeat them, a Lookback feature was proposed for \textit{the Lookback RNN}~\cite{magenta2016}.
A user study suggested that the RNN with Lookback feature outperforms basic RNN~\cite{yang2017midinet} so we also use it in our model\footnote{For fair comparison in experiments, all models were equipped with this feature.}.
The lookback distance is 2 and 4 for the Bar Layer, 4 and 8 for the Beat Layer, 4 and 8 for the Note Layer.
Therefore, the Note Layer without the condition of the Beat layer and Bar layer is equivalent to \textit{the Lookback RNN}.

\subsection{LSTM-Based Event Sequence Generator}
Bar profiles, beat profiles and notes can be abstracted as events, which can be generated by RNN.
It might be better to use different models for generating different types of events, but for simplicity we use the same LSTM-based event sequence generator for the Bar Layer, Beat Layer and Note Layer.

The event sequence generator $G_\theta$ is trained by solving the following optimization problem:
\begin{equation}\label{equ:condition}
\max_\theta \sum_{y \in \mathcal{Y}} \sum_{t=1}^{\text{len}(y)} \log p(y_t|y_0,...,y_{t-1}, c_t)
\end{equation}
where $\theta$ are the parameters of the generator, $y$ is a sequence sampled in the event sequences dataset $\mathcal{Y}$.
And $y_t$ denotes the $t$-th event in $y$, $c_t$ denotes the condition for $y_t$.

LSTM is used to predict the conditional probability in Eq.~\eqref{equ:condition}, which is characterized by input gates $i_t$, output gates $o_t$ and forgetting gates $f_t$~\cite{hochreiter1997long}:
\begin{equation}
\begin{aligned}
i_t & = \sigma(W_{ix}x_t + W_{im}m_{t-1}) \\
f_t & = \sigma(W_{fx}x_t + W_{fm}m_{t-1}) \\
o_t & = \sigma(W_{ox}x_t + W_{om}m_{t-1}) \\
c_t & = f_t \odot c_{t-1} + i_t \odot \tanh(W_{cx} x_t + W_{cm} m_{t-1}) \\
m_t & = o_t \odot c_t \\
p_t & = \text{Softmax}(m_t)
\end{aligned}
\end{equation}
where $W_{ix}$, $W_{im}$, $W_{fm}$, $W_{fm}$, $W_{ox}$ and $W_{om}$ are trainable parameters, $\odot$ denotes the element-wise multiplication and $\sigma(\cdot)$ denotes the sigmoid function.
The $y_{t-1}$ is used as input $x_t$.

The lookback feature is added to the Bar Layer, Beat Layer and Note Layer, to help the model memorize recent events and potentially repeat them. The lookback distance is 2 and 4 for the Bar Layer, 4 and 8 for the Beat Layer, 4 and 8 for the Note Layer.

During generation, given a primer sequence as an initial input sequence, the LSTM network generates the distribution $p_0$ over all candidate events.
The next event was chosen by sampling over $p_0$.
The successive events are generated according to $p(y_t|y_0,...,y_{t-1}, c_t)$.

\section{Experiments}
\label{Sec:Exp}

Evaluating the performance of the models for melody generation is difficult.
The main reason is that measuring the quality of the generated melodies is subjective and it is hard to find an objective metric.
% Carrying out human behavioral experiments is often difficult and expensive.
% Another reason is that currently, most models about this task is not open-sourced, which makes it hard to compare different models.

We evaluated three generative models, HRNN-1L, HRNN-2L and HRNN-3L mainly based on behavioral experiments.
HRNN-3L is the model we described in the previous section.
HRNN-2L is the HRNN-3L without the Bar Layer while HRNN-1L is the HRNN-3L without the Bar Layer and the Beat Layer.
Note that HRNN-1L is actually the \textit{Lookback RNN} developed by Google Magenta \cite{magenta2016}.
The music pieces generated by the models were not post-processed.

% We compared HRNN-3L with two state-of-the-art melody generation models MidiNet~\cite{yang2017midinet} and MusicVAE~\cite{pmlr-v80-roberts18a} in the experiment described in section \ref{subsec:com_midinet}.

All melodies used in experiments were publicly available \footnote{\url{https://www.dropbox.com/s/vnd6hoq9olrpb5g/SM.zip?dl=0}}.

\subsection{Implementation Details}
\begin{figure}
\centering
\includegraphics[width=\linewidth]{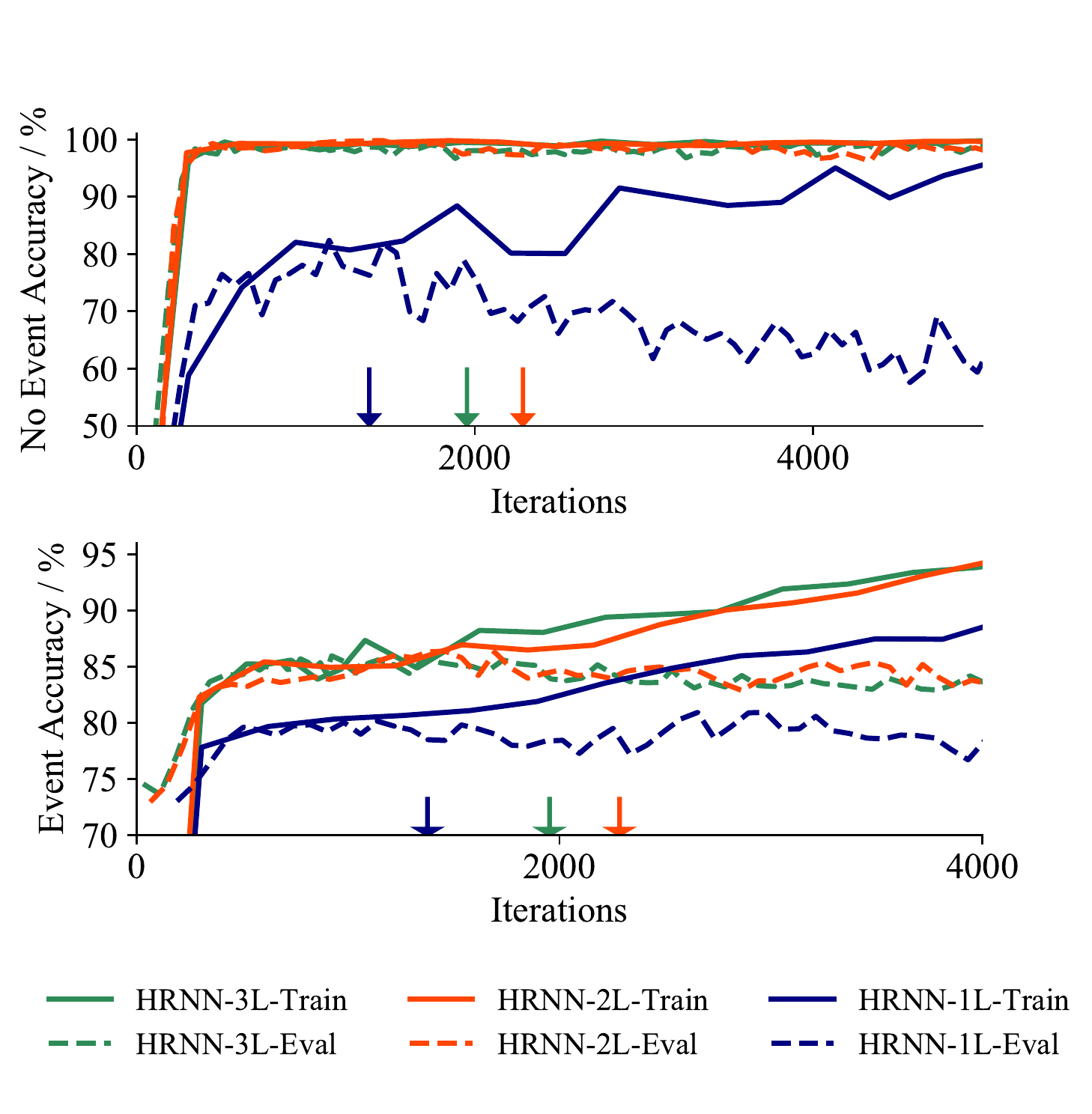}
\caption{The accuracy curves of Note Layer for training and validation dataset. \textbf{Left}: no-event accuracy curves. \textbf{Right}:  event~(both note-on event and note-off event) accuracy curves. Arrows indicate iterations at which training stopped to prevent over-fitting.}
\label{Fig:Accuracy}
\end{figure}

All LSTM networks used in experiments had two hidden layers and each hidden layer had 256 hidden neurons.
They were trained with Adam algorithm~\cite{kingma2014adam} and the initial learning rate was 0.001.
The minibatch size was 64.
The $\beta_1$, $\beta_2$ and $\epsilon$ of Adam optimizer were set to 0.9, 0.999, 1e-8.
To avoid over-fitting, dropout with ratio 0.5 was adopted for every hidden layer of LSTM and validation-based early stopping~(see Fig.~\ref{Fig:Accuracy}) was employed so that the training was stopped as soon as the loss on the validation set increased for 5 times in a row~(the model is evaluated on validation set every 20 training iterations).

In each generation trial, primer sequences~(both profiles and melodies) were randomly picked from the validation dataset.
For the Bar Layer and Beat Layer, one profile is given as the primer.
For the Note Layer, the length of the primer sequence is 1 beat.
% A frequently used method for sequence generation is Beam Search, which iteratively generates next event with the k best event sequences at time t and keeps the resulting k best sequences at time t + 1.
Beam search with a beam of size 3 was used in all experiments.

\subsection{Dataset}\label{SubSec:Dataset}
We collected 3,859 lead sheets with the time signature of 4/4 in MusicXML format from \url{http://www.wikifonia.org}. We have made these lead sheets publicly available\footnote{\url{https://www.dropbox.com/s/x5yc5cwjcx2zuvf/Wikifonia.zip?dl=0}}
% \footnote{To avoid violation of the ``double-blind'' review policy, the website is not indicated in this version of the paper.}.
90\% of the lead sheets were used as training set and the other 10\% were used as validation set.
% MusicXML files saved melody and chord information separately and thus make it easier and more accurate to extract melody and chord.
The speed of most music pieces in the dataset is 120 beats per minute.
To guarantee the correct segmentation of melodies, all melodies started with weak beats were removed so that we can take bar as a basic unit.

\subsection{Guiding Effect of Profiles}

To verify whether the beat and bar profiles can guide the generation of melody, we plotted the Note Layer's accuracy curves of in Fig.~\ref{Fig:Accuracy}.
Here both event accuracy~(accuracy of the note-on event and note-off event; chance level is $1/ 37$) and no-event accuracy~(accuracy of the no-event; chance level is $1/2$) are plotted.

With beat and bar profiles, the Note Layer learned the pattern of no-event quickly and easily.
For models with profiles, the accuracy of no-event increased to nearly 100\% at about 200 iterations while the model without profile converged slowly and over-fitting started after about 2000 iterations.
Since rhythm is encoded by no-event~(see Section~\ref{Sec:MelodyRepresentation}), this showed that the Note Layer successfully utilized the rhythm provided by the beat and bar profiles.
The accuracy of note-on and note-off events also improved,
which means models with profiles not only did a good job in predicting rhythm, but also in predicting pitch.

\begin{figure}
\centering
Beat profiles: "1, 1, 1, 2, ..., 1, 1, 1, 2"{
\includegraphics[width=0.48\textwidth]{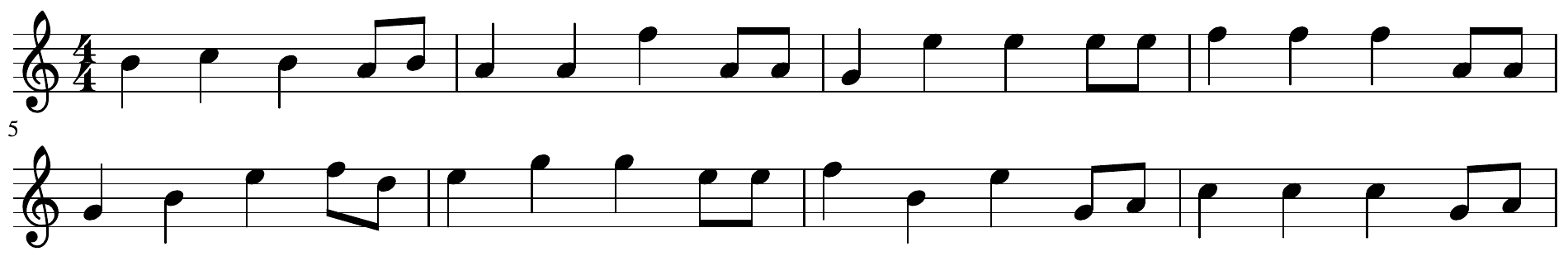}
\label{Fig:beat_profile_44488}
} \\
Beat profiles: "2, 1, 2, 1, ..., 2, 1, 2, 1"{
\includegraphics[width=0.48\textwidth]{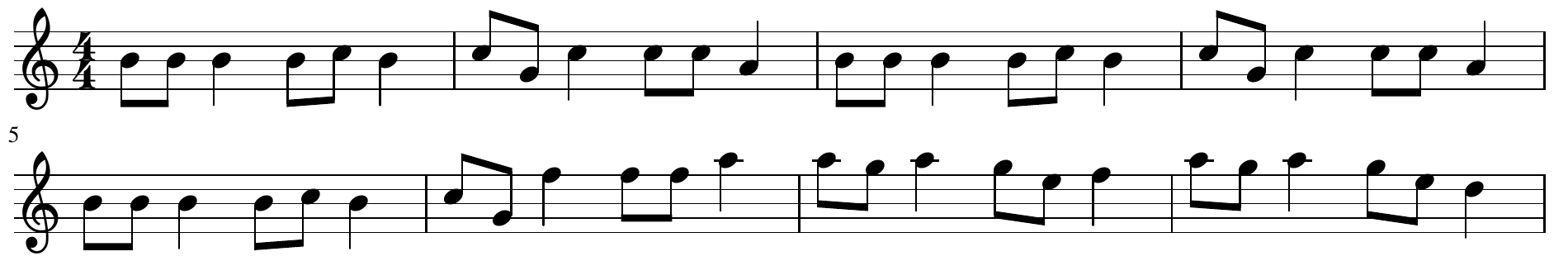}
\label{Fig:beat_profile_2121}
} \\
Beat profiles: "2, 5, 2, 5, ..., 2, 5, 2, 5"{
\includegraphics[width=0.48\textwidth]{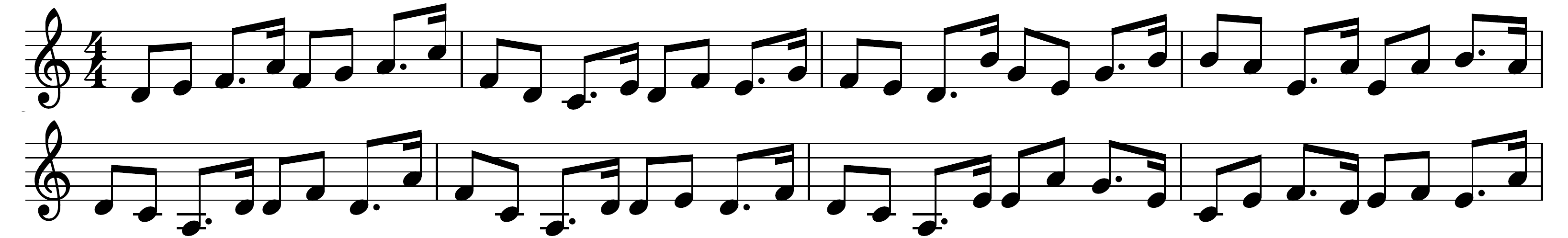}
\label{Fig:beat_profile_2525}
}
\caption{Melodies generated with given beat profile sequences.}
\label{Fig:beat_profile}
\end{figure}

\begin{figure}
\centering
Bar profiles: "1, 2, 1, 2, 1, 2, 1, 2"{
\includegraphics[width=0.48\textwidth]{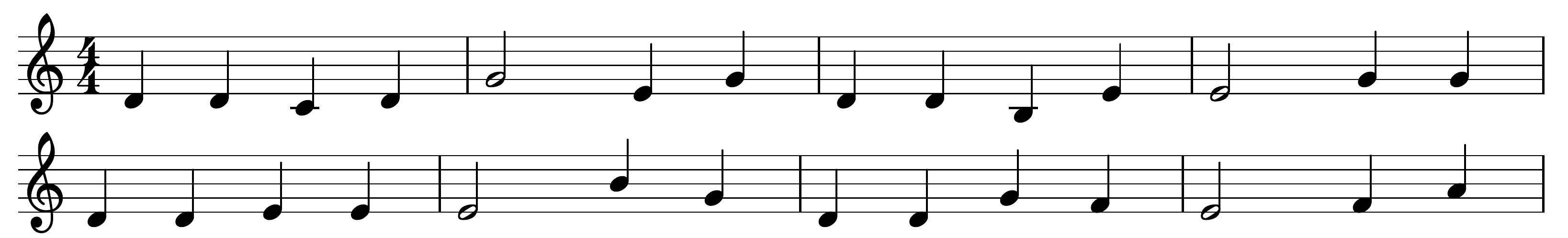}
} \\
Bar profiles: "2, 3, 2, 3, 2, 3, 2, 3"{
\includegraphics[width=0.48\textwidth]{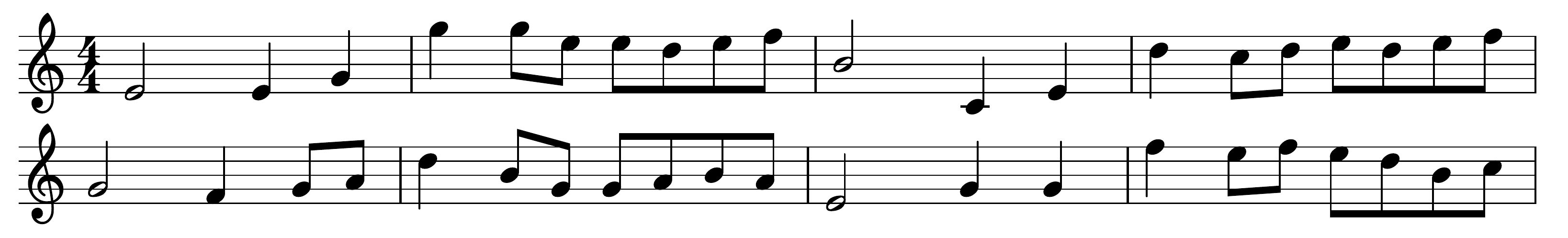}
} \\
Bar profiles: "3, 1, 3, 1, 3, 1, 3, 1"{
\includegraphics[width=0.48\textwidth]{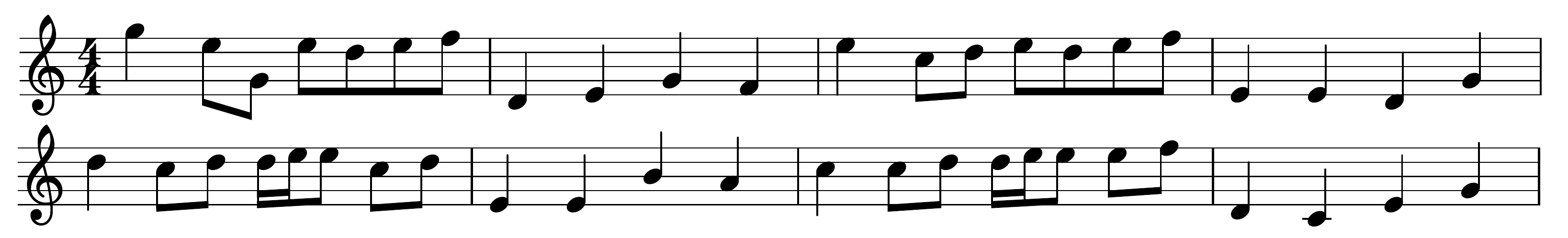}
}
\caption{Melodies generated with given bar profile sequences.}
\label{Fig:bar_profile}
\end{figure}

With given profile sequences, the Note Layer will generate melodies with rhythm represented by profile sequence.
To show this, we used handcrafted profile sequences to guide the generation of the Note Layer.
Fig.~\ref{Fig:beat_profile} and Fig.~\ref{Fig:bar_profile} show generated melodies given beat profile sequences to HRNN-2L and bar profile sequences to HRNN-3L~(profile index in Fig.~\ref{Fig:Profiles}).
The results verified that the generated melodies are strongly constrained by the given profile sequence patterns.
The same conclusion can be obtained using fixed beat profiles and bar profiles extracted from existing melodies.
We extracted beat profiles and bar profiles of children's rhymes \textit{Twinkle, Twinkle Little Star} and generated melody conditioned on these profiles.
The result is shown in Fig.~\ref{Fig:Twinkle}. The audio files can be found in \textbf{Supplementary Materials}. The rhythm of the generated melody is unison with the original melody, which suggests that the beat profiles and bar profiles effectively guided the generation of melody.

\begin{figure}
\centering
Original melody{
\includegraphics[width=0.48\textwidth]{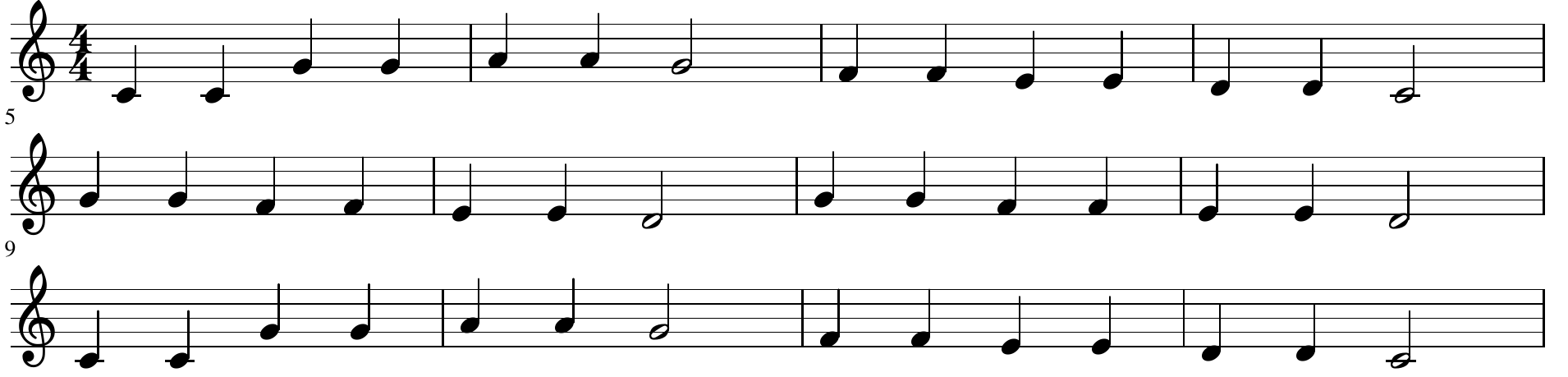}
} \\
Generated melody{
\includegraphics[width=0.48\textwidth]{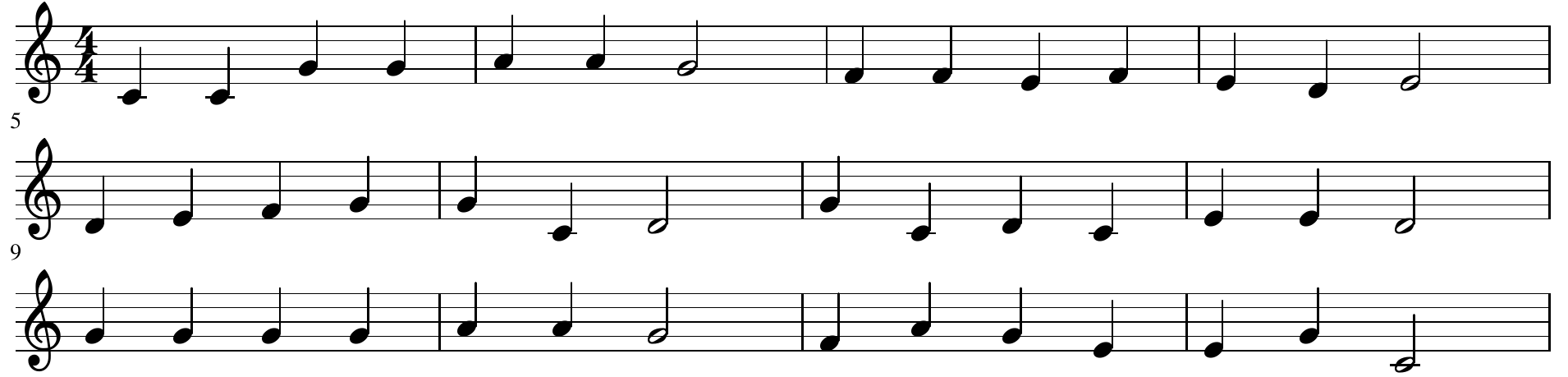}
}
\caption{Original melody of Twinkle, Twinkle Little Star and  generated melody by Note Layer of HRNN-3L, given profiles of the original melody.}
\label{Fig:Twinkle}
\end{figure}

% We generated beat profile sequences with the Beat Layer conditioned on handcrafted bar profile sequences and then generated melodies conditioned on these generated beat profile sequences to visualize the influence of bar profiles to beat profiles.

% \begin{figure}
% \centering

% \begin{minipage}[b]{1.0\linewidth}
% 	\centering
%     \centerline{\epsfig{figure=Figs/twinkle_origin.pdf,width=8.5cm}}
%     \centerline{Original melody.}
% \end{minipage}

% \begin{minipage}[b]{1.0\linewidth}
% 	\centering
%     \centerline{\epsfig{figure=Figs/twinkle_hrnn.pdf,width=8.5cm}}
%     \centerline{Generated melody.}
% \end{minipage}

% \caption{Original melody of Twinkle, Twinkle Little Star and  generated melody by Note Layer of HRNN-3L, given profiles of original melody.}
% \label{Fig:Twinkle}
% \end{figure}

% The same conclusion can be obtained using fixed beat profiles and bar profiles extracted from existing melodies.
% We extracted beat profiles and bar profiles of children's rhymes \textit{Twinkle, Twinkle Little Star} and generated melody conditioned on these profiles (the first bar of the original melody is used to prime the model).
% Fig.~\ref{Fig:Twinkle} illustrates the original melody and the melody generated by the Note Layer given its extracted profiles.
% The generated melody is provided in the supplementary material.
% The rhythm of the generated melody is unison with the original rhythm.
% This means that the beat profiles and bar profiles effectively guided the generation of melody.

\subsection{Qualitative  Comparison}

\begin{figure*}
\centering
\includegraphics[width=\textwidth]{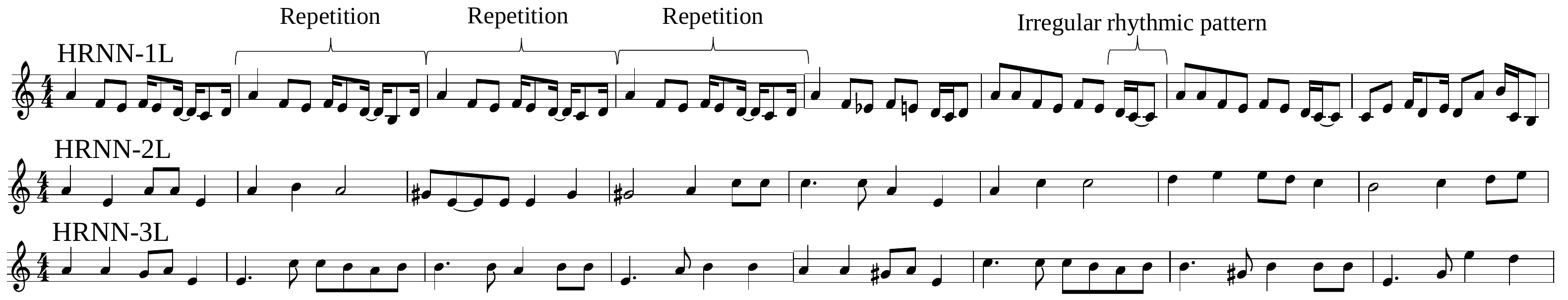}
\caption{Melodies generated by HRNN-1L, HRNN-2L and HRNN-3L.
}
\label{Fig:123l}
\end{figure*}

% The above results have shown the strong guiding effect of the Bar Layer and Beat Layer for the Note Layer.
The strong guiding effect of profiles implies that the Note Layer could output good melodies if higher layers could generate good profile sequences.
Since note sequences are much longer than their profile sequences, learning the latter should be easier than learning the former using the same type of model.
% Also, these profiles could help generate rhythmic pattern more accurately when used as a condition to generate melody.
Thus, compared to HRNN-1L, melodies generated by HRNN-2L and 3L model should be more well-organized and keep better long-term structures.
The qualitative comparison verified this point.
Three typical music pieces generated by HRNN-1L, HRNN-2L, HRNN-3L with the same primer note were shown in Fig.~\ref{Fig:123l}.
The melody generated by HRNN-1L has basic rhythm, but also irregular rhythmic patterns.
And the melodies generated by HRNN-2L and HRNN-3L contain less irregular rhythmic patterns. %, and their period is more of variety rather than simple repetition.

%Melodies generated by HRNN-2L and HRNN-3L model are more well-organized because HRNN-2L and HRNN-3L have the bar and beat profile.
%These profiles could help generate rhythmic pattern more accurately, when used as a condition to generate melody.
%And in the long term structure,
% And in the long term structure, since HRNN-2L and HRNN-3L have Lookback feature on the higher-level-Layer (Bar Layer and Beat Layer),  the profile maybe repeated periodically,
%Since the profile is only a condition when generating melody, the melody itself will still have some variety (both in pitch and rhythm) in the Note Layer.
%Thus, compared to HRNN-1L, melody by HRNN-2L and 3L model would have not repetition but the phrase in the long-term structure.

\subsection{Comparison of Different Number of Layers}
Three human behavior experiments were conducted to evaluate melodies generated by models.
For this propose, we built an on-line website where people could listen to melodies and give their feedback.
To model real piano playing scenario, sustain pedal effect was added to all model generated and human composed musical pieces evaluated in these experiments. This was achieved by extending all notes' duration so that they ended at the end of the corresponding bars.

\subsubsection{Two-Alternative Forced Choice Experiment}
\label{subsec:AFC}
% To evaluate these melody generative models, we employed two-alternative forced choice (2AFC)  to measure the influence given by HRNN architecture.
We randomly provided subjects pairs of melodies with the length of 16 bars (about 32 seconds) and asked them to vote  (press one of two buttons in the experiment interface) which melody sounded better in every pair.
This is the two-alternative forced choice~(2AFC) setting. Subjects had infinite time for pressing the buttons after they heard the melodies. Pressing the button started a new trial.

Three types of pairs were compared:  HRNN-3L versus HRNN-1L, HRNN-2L versus  HRNN-1L and HRNN-3L versus HRNN-2L.
Each model generated a set of 15 melodies and in every trial two melodies were randomly sampled from the two corresponding sets.
Different types of pairs were mixed and randomized in the experiment. 
% Sample audio files of melodies generated by the three models can be found in \textbf{Supplementary Materials} (in the directory \textsf{melody}).

\begin{figure}
\centering
\includegraphics[width=\linewidth]{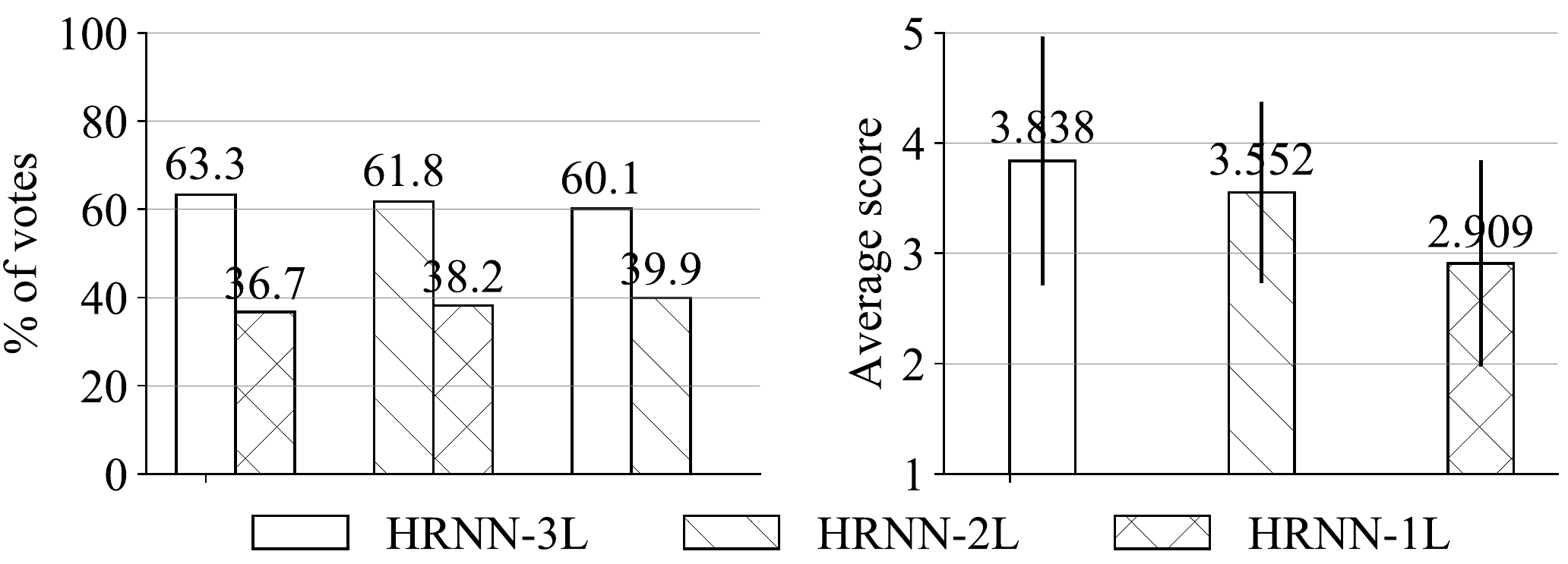}
\caption{Results of the 2AFC experiment~(\textbf{left}) and the melody score experiment~(\textbf{right}).}
\label{Fig:Exp}
\end{figure}

Call for participants advertisement was spread in a social media.
1637 trials were collected from 103 IP addresses~(Note that one IP address may not necessarily correspond to one subject).
The results are shown in Fig.~\ref{Fig:Exp}.
In nearly two-thirds of trials, melodies generated by hierarchical models were favored (Pearson's chi-squared test, $p=3.96\times10^{-10}$ for HRNN-3L versus HRNN-1L and $p=2.70\times10^{-8}$ for HRNN-2L versus HRNN-1L).
In addition, subjects voted more for melodies generated by HRNN-3L than by HRNN-2L~($p=3.38\times10^{-6}$)
% which proves the improvement given by HRNN architecture.

\begin{figure}
\centering
\includegraphics[width=0.7\linewidth]{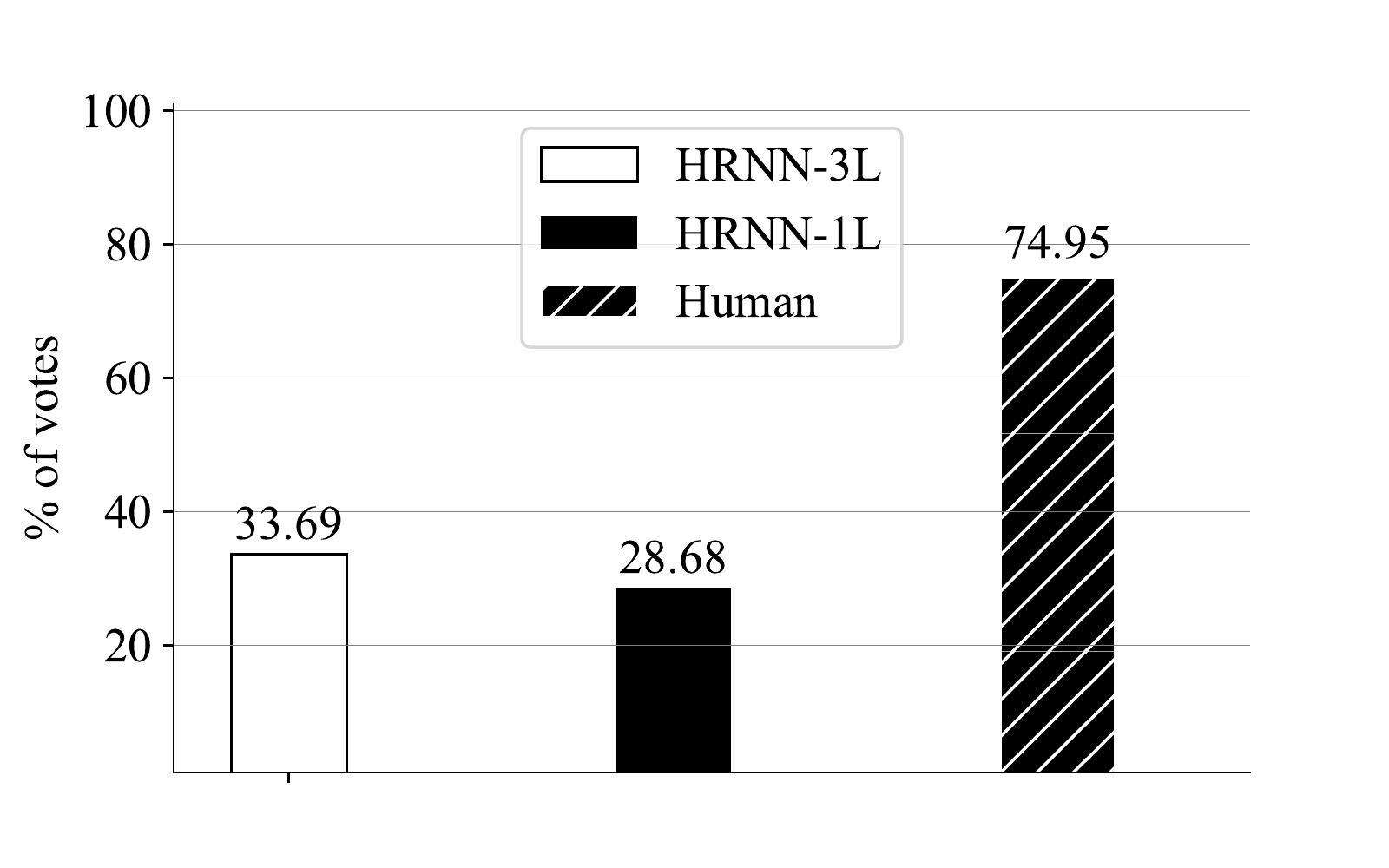}
\caption{Results of the music turing test.}
\label{Fig:turing}
\end{figure}

\subsubsection{Melody Score Experiment}
To quantitatively measure the quality of melodies generated by different models and verify the conclusion obtained in the online experiment,  we invited 18 subjects between ages of 18 and 25 to score these melodies.

Every subject was asked to score every melody used in 2AFC experiment with 5 levels: 5 the best and 1 the worst.  It took each subject about 24 minutes to finish the experiment.

We calculated the average score of every melody in Fig.~\ref{Fig:Exp}.
The results verified that the two additional layers improved the quality of melodies generated by the single-layer model~(two-tailed test, $p = 0.00576$ for HRNN-3L versus HRNN-1L).

\subsubsection{Control the Number of Parameters}
In the above experiment, the number of parameters of HRNN-3L (2.84M) was three times that of HRNN-1L (0.94M). 
That might be the reason why HRNN-3L performed better. 
So we trained a HRNN-1L model with 450 hidden neurons~(2.79M) and conducted another 2AFC experiment~(as described in Section 5.5.1) to compare their performances. 
A total of 203 trials were collected from 21 IP addresses. In 127 trials (62.6\%, $p = 3.44\times 10^{-4}$), melodies generated by HRNN-3L were favored, which is similar to the result (63.3\%) in comparison with HRNN-1L with fewer parameters~(Fig.~\ref{Fig:Exp} left). 
The results indicate that the better performance of the hierarchical structure was not mainly due to the increased number of parameters. 

\subsubsection{Music Turing Test}
To compare the quality of the melodies generated by the models and the quality of melodies composed by human composers, a music ``Turing test" was carried out.
Only two models, HRNN-3L and HRNN-1L, were tested.
% Since human music contained melody and chord, chords were added as a condition of the Beat Layer and Note Layer.
We found that without chord, it was difficult for the models to generate melodies that could fool human.
So chords were added as a condition of the Beat Layer and Note Layer in training and generation.
Chords and primer sequences used in the generation of a melody were extracted from the same music piece in the validation set.

A total of 50 musical pieces containing 8 bars were randomly chosen from the validation set as human composed music.
Then the HRNN-1L and the HRNN-3L both generated 25 melodies with the length of 8 bars. Sample % audio files of human composed musical pieces and audio files generated by HRNN-1L and the HRNN-3L can be found in \textbf{Supplementary Materials} (in the directory \textsf{melody\_chord}).
We provided subjects music pieces from these 100 examples and asked them to judge if they were composed by human by pressing one of two buttons in the experiment interface. Subjects had infinite time for pressing the buttons after they heard the melodies. Pressing the button started a new trial.
% Whether if they did the right choice is provided once they finished a trial.
Feedback about the correctness of the choice was provided immediately after the subjects made the choice in every trial.
Then the subjects had a chance to learn to distinguish human-composed and machine-composed melodies, which made it hard for the models to pass the Turing Test.

In this experiment, we collected 4185 trials from 659 IP addresses, among which 1018 music pieces were generated by HRNN-1L, 1003 by HRNN-3L and 2164 by human. As shown in Fig.~\ref{Fig:turing}, 33.69\% of music pieces generated by HRNN-3L were thought to be human composed (or real), which is higher than the result of HRNN-1L, 28.68\%.

It is seen that not all music pieces sampled from the original dataset were thought to be composed by humans (only 74.95\% were correctly classified).
This implies that some music pieces generated by the models sounded better than human composed pieces,
and that the quality of the dataset is not very high.
% If we normalize the scores such that all human composed pieces are correctly identified, then 33.69/74.95=44.94\% of generated pieces confuse the subjects.

\subsection{Comparison with Other Models}
\label{subsec:com_midinet}

Though many models have been proposed for melody generation, to the best of our knowledge, only the lookback RNN [21], attention RNN [21], MidiNet [23] and MusicVAE [19] have public available source codes. 
These models represent the state-of-the-art in the area of melody generation. 
It was reported in~\cite{yang2017midinet} that the attention RNN had similar performance to the lookback RNN.
Our previous experiments have shown that HRNN-3L performed better than the lookback RNN, i.e. HRNN-1L.
We then compared MidiNet and MusicVAE with HRNN-3L based on human evaluations.

\subsubsection{MusicVAE}

MusicVAE is a variatiaonal autoencoder that can generate melodies with the length of 16 bars.
We compared the HRNN-3L model with a MusicVAE trained on our dataset~(with the same training setting in the original paper) and their pretrained MusicVAE using the 2AFC setting separately.
Each model generated 15 melodies with the length of 16 bars.
In each experiment, we randomly provided subjects 20 pairs of melodies and subjects were asked to vote for the better sounded melody.

In the comparison between HRNN-3L and MusicVAE that was trained on our dataset, 435 trials were collected from 17 IP addresses.
In  317 trials~(72.6\%,  $p =1.41 \times 10^{-21}$), melodies generated by HRNN-3L were favored.
We found the rhythm of melody generated by this MusicVAE is chaotic.
One reason might be that the size of our dataset is too small compared with the dataset used in~\cite{pmlr-v80-roberts18a}.

We then compared HRNN-3L and Pretrained-MusicVAE.
 461 trials were collected from  21 IP addresses. In 293 trials~(63.5\%, $p =5.82 \times 10^{-9}$), melodies generated by HRNN-3L were favored.
We generated 200 melodies with the Pretrained-MusicVAE and the statistics on 200 melodies show that about 40\% of notes generated by Pretrained-MusicVAE had pitches lower than C2, which made some melodies sound strange.

\subsubsection{MidiNet}
Another 2AFC experiment was used to compare HRNN-3L with MidiNet~\cite{yang2017midinet}. 
The MidiNet was trained on our dataset with the same training setting in the original paper.
Since MidiNet required chords as an input, chords were used as a condition for MidiNet and HRNN-3L.
MidiNet generated 25 melodies with the length of 8 bars conditioned on chords. 
The 25 melodies of HRNN-3L used in the Music Turing Test were used here for comparison..
In this 2AFC experiment, we randomly provided subjects pairs of melodies (HRNN-3L versus MidiNet) and asked them to vote for the better sounded melody in every pair. 
290 trials were collected from 28 IP addresses. 
In 226 trials (77.9\%, $p =1.85 \times 10^{-21}$), melodies generated by HRNN-3L were favored.

\section{Discussions}
\label{Sec:Con}
In this paper, we present a hierarchical RNN model to generate melodies.
Two high-level rhythmic features, beat profile and bar profile, are designed to represent rhythm at two different time scales respectively.
The human behavior experiment results show that the proposed HRNN can generate more well-organized melodies than the single-layer model.
In addition, the proposed HRNN, though very simple, can generate better melodies than the well-known models MusicVAE and MidiNet. 

In the Music Turing Test, only 33.69\% pieces generated by the proposed model were thought to be composed by human.
This proportion is still far from our expectation and there is still a long way to go for developing a perfect automatic melody generator.
However, under current technology, HRNN has achieved good enough results.
On one hand, one should notice that automatic generation of other forms of data is in the similar stage. For example, many state-of-the-art machine learning models trained on natural images~\cite{zhu2017unpaired}\cite{isola2016image} generate no more than 30\% images that can fool human.
On the other hand, the dataset used in this study is not good enough (only 74.95\% human composed pieces were thought to be composed by human) which has hurt the performance of the model.
If the model is trained on a dataset in which nearly all human composed pieces can be correctly classified, one may expect that about  44.9\% (=33.69/74.95) pieces generated by the model would fool human subjects.

% There are a number of limitations with the current model. First,
% the dataset only contains musical pieces with 4/4 time signature.
% More time signatures should be taken into consideration to improve the power of the model.
% Second, since we quantized a bar into 16 time step, the encoding could not represent triplet or other types of rhythm.
% Third, the profiles presented in this paper are only about beats and bars.
% An appropriate representation for melody's features at larger time scales is demanded.

The proposed approach of course have many limitations which should be considered in future. First, since we quantized a bar into 16 time step, the encoding could not
represent triplet or other types of rhythm. Second, we only selected musical pieces with 4/4 time signature from the original dataset for training. More time signatures should be taken into consideration
to improve the capability of the model. Third, we only considered beats and bars as larger units than notes, and did not consider phrases which are often present in pop music, since they are not labeled in the dataset. With larger time scale units, the model may output pieces with longer temporal structure.

\section*{Acknowledgements}
This work was supported in part by the National Natural Science Foundation
of China under Grant Nos. 61332007, 61621136008 and 61620106010.

\bibliographystyle{aaai}
\bibliography{icme2018}

\end{document}